# Rational Strain Engineering in Delafossite Oxides for Highly Efficient Hydrogen Evolution Catalysis in Acidic Media

Filip Podjaski[1,2], Daniel Weber[1,3], Siyuan Zhang[4], Leo Diehl[1,3], Roland Eger[1], Viola Duppel[1], Esther Alarcon-Llado[5], Gunther Richter[6], Frederik Haase[1,3], Anna Fontcuberta i Morral[2,7], Christina Scheu[4], Bettina V. Lotsch*[1,3,8,9]

[1]Max-Planck-Institute for Solid State Research, Heisenbergstraße 1, 70569 Stuttgart, Germany. [2]Laboratory of Semiconductor Materials, Institute of Materials, Faculty of Engineering, Ecole Polytechnique Fédérale de Lausanne, Station 12, 1015 Lausanne, Switzerland. [3]Department of Chemistry, University of Munich (LMU), Butenandtstraße 5-13, 81377 München, Germany. [4]Max-Planck-Institut für Eisenforschung GmbH, Max-Planck-Straße 1, 40237 Düsseldorf, Germany. [5]AMOLF, Science Park 104, 1098 XG Amsterdam, The Netherlands. [6]Max-Planck-Institute for Intelligent Systems, Heisenbergstr. 3, 70569 Stuttgart, Germany. [7]Institute of Physics, Faculty of Basic Sciences, EPFL, 1015 Lausanne, Switzerland. [8]Nanosystems Initiative Munich (NIM), Schellingstraße 4, 80799 München, Germany. [9]Center for Nanoscience, Schellingstraße 4, 80799 München, Germany.

*Corresponding author: b.lotsch@fkf.mpg.de

# Abstract

The rational design of hydrogen evolution reaction (HER) electrocatalysts which are competitive with platinum is an outstanding challenge to make power-to-gas technologies economically viable. Here, we introduce the delafossites $PdCrO_2$, $PdCoO_2$ and $PtCoO_2$ as a new family of electrocatalysts for the HER in acidic media. We show that in $PdCoO_2$ the inherently strained Pd metal sublattice acts as a pseudomorphic template for the growth of a strained (by +2.3%) Pd rich capping layer under reductive conditions. The surface modification continuously improves the electrocatalytic activity by simultaneously increasing the exchange current density $j_0$ from 2 to 5 mA/cm²$_{geo}$ and by reducing the Tafel slope down to 38 mV/decade, leading to overpotentials $\eta_{10}$< 15 mV for 10 mA/cm²$_{geo}$, superior to bulk platinum. The greatly improved activity is attributed to the in-situ stabilization of a $\beta$-palladium hydride phase with drastically enhanced surface catalytic properties with respect to pure or nanostructured palladium. These findings illustrate how *operando* induced electrodissolution can be used as a top-down design concept for rational surface and property engineering through the strain-stabilized formation of catalytically active phases.

# Introduction

Global warming and the decreasing availability of fossil fuels urge today's society to transition to more sustainable energy sources. While there is enough solar and wind power to satisfy our needs in terms of total energy, [1, 2] the available power fluctuates strongly and requires intermediate and long term storage. [3, 4] One viable option is the storage of the intermittent electrical energy in the form of chemical fuels such as hydrogen (power-to-X). Clean hydrogen can be produced by alkaline electrolyzers, which require constant and high current densities for a stable operation.[5] In acidic environments, more powerful and flexible proton exchange membrane (PEM) electrolyzers can be employed that allow for coupling with fluctuating energy sources such as wind and solar. [6, 7] As the most widely used electrocatalysts for the hydrogen evolution reaction (HER) is still platinum, an expensive and scarce material that is also poisoned easily, research into alternative or modified highly efficient and stable electrocatalysts under various conditions has been identified as a key goal in energy science.[8, 9, 10]

Strain effects have been discussed to be at the heart of enhanced intrinsic activities toward several catalytic reactions including the HER, as predicted and observed for example in Pd overlayers. [11, 12, 13] More recently, the scope of this concept has been widened and the direct strain control in substrate induced strain effects or in core-shell particles have been effectively used for the HER as well as the oxygen evolution reaction (OER) and the oxygen reduction reaction (ORR). [14, 15, 16, 17, 18, 19]

Here, we report the time evolution of the electrocatalytic activity of the $ABO_2$ delafossites $PdCrO_2$, $PdCoO_2$, and $PtCoO_2$ for the HER in acidic medium and show how inherent structural strain can be used to enhance the catalytic efficiency *operando*. Since their discovery and the observation of their unusually high and anisotropic conductivity in 1971, [20, 21, 22] these oxides have attracted renewed interest recently for their unusual electronic properties, [23, 24] anisotropic thermopower,[25] and most recently, for the discovery of hydrodynamic electron flow occurring in nanostructured $PdCoO_2$.[26]

A common property of the metallic Pd and Pt based delafossites is an inherently expanded hexagonal metal sublattice with extended nearest neighbour distances on the A site compared to the pure metals (2.830 Å in $PdCoO_2$ and 2.923 Å in $PdCrO_2$ vs. 2.751 Å in *fcc* Pd(111); 2.823 Å in $PtCoO_2$ vs. 2.775 Å in metallic Pt(111)). [20, 21, 22, 24] This sublattice is separated by a layer of edge-sharing $MO_6$ (M = Co, Cr) octahedra, see Fig. 1 a, which gives rise to anisotropic transport properties. So far, these and other delafossite oxides have been reported as highly efficient electrocatalysts for the OER in alkaline media. [27, 28, 29, 30] Furthermore, copper- and silver-based delafossites have been reported as photocathodes due to their large band gap.[31, 32, 33, 34] Apart from $AgRhO_2$ and $CuCrO_2$, the stability of these systems appears to be limited to basic, neutral or non-reductive conditions. [35, 36, 37]

In this work, we make use of the intrinsically strained metal sublattices and study the hydrogen evolution activity of the delafossite oxides $PdMO_2$ (M = Cr, Co) and $PtCoO_2$ for the first time, investigating the influence of strain on the catalytic activity. For $PdCoO_2$, our long-term studies reveal a gradual enhancement of the already excellent water reduction activity of the bulk material over time, putting the electrocatalytic activity of *operando* modified $PdCoO_2$ *en par* with that of the top-of-the-class HER electrocatalyst in acidic medium, platinum. We elucidate how the charge transfer properties as well as the high specific activity per surface area evolve, which we attribute to the strain-induced stabilization of *in-situ* formed, catalytically highly active β-palladium hydride ($PdH_x$, x~0.62-0.67).[38]

Together with an apparent increase in exchange current density, these effects lead to a drastically reduced overpotential at 10 mA/cm²$_{geo}$ ($\eta_{10}$) < 15 mV and a Tafel slope of 30-40 mV/decade, enabling stable operation at current densities of 100 mA/cm²$_{geo}$ with less than 100 mV of applied (uncorrected) potential.

## Results

**Electrochemical characterization.** Polycrystalline powders of the delafossites $PdCrO_2$, $PdCoO_2$ and $PtCoO_2$ were prepared according to previously reported procedures (for details, see the Method section). The crystal structure of the isostructural compounds (space group *R*-3*m*) was confirmed by powder XRD measurements (see Fig. S1) and is schematically displayed for $PdCoO_2$ in Fig. 1 a.[20, 39, 40, 41] The crystallite sizes were in the range from 1-3 μm for $PdCrO_2$, 10-30 μm for $PtCoO_2$ and up to 1 mm for $PdCoO_2$ as evidenced by electron microscopy (see Fig. 2). These microcrystals were subsequently pressed into a carbon paste electrode (see Methods for details) to study their activity towards HER in acidic media. The analysis was performed in hydrogen saturated 1M $H_2SO_4$ by means of cyclic voltammetry (CV), chronopotentiometry and –amperometry. The polarization curves in Fig. 1 b-d show the uncorrected cathodic currents per geometric surface area (in $mA/cm^2_{geo}$) for the first 1000 cycles on all three materials, which all increase in efficiency in different ways. After correcting for the series resistance losses (*IR*-drop) extracted from impedance measurements before the respective cycles, the CVs can be fitted to the Tafel equation

$$\eta = b \log(i/j_0)$$

with $\eta$ being the overpotential with respect to the reversible hydrogen electrode (RHE), while *b* denotes the Tafel slope (in mV/decade), which provides insight onto the electrocatalytic processes occurring on the surface, $i$ being the current density and $j_0$ the exchange current density, which is a measure of the intrinsic activity per surface area.[42]

The activity of all delafossites is very high, requiring an overpotential $\eta_{10}$ of far less than 100 mV for all systems. $PdCrO_2$ initially requires $\eta_{10}$ of approx. 50 mV (Fig. 1 b and e). Subsequently, the current density decreases and stabilizes with increasing cycle number. $PdCoO_2$ (Fig. 1 c and e) initially also yields $\eta_{10}$ at 50 mV vs. RHE for a similar loading and keeps improving for the first 1000 cycles down to 12(3) mV. In contrast, PtCoO$_2$ initially shows a lower activity than the Pd based materials and quickly improves towards comparable activities of PdCoO$_2$ within 30-100 cycles. Afterwards, it stabilizes at $\eta_{10}$ around 30-35 mV. Remarkably, the overpotentials of all the materials are very low after some initial cycles and especially the Co based delafossites outperform most catalysts already after 30 cycles ($\eta_{10}$ < 40mV), particularly all oxides, in acidic media. [43, 44, 45, 46]

To better understand the curious evolution of the overpotential over time, we discriminate between the effects arising from the intrinsic activity and those induced *operando*, i.e. during the electrocatalytic process. In the first cycles, the cathodic currents are influenced by surface activation or hydrogen sorption, which is well known for Pt and Pd, respectively. [47, 48] This is clearly visible in the first cycle of PdCrO$_2$ (Fig. 1 b), where the apparent high activity occurs already at the RHE potential before hydrogen saturation at the surface is achieved. In the range of 10-1000 cycles, the catalytic mechanism on the surface evolves with increasing cycle number and the currents tend to stabilize. Interestingly, the overpotential is modified also by the catalytic process itself, which is reflected by the decreasing Tafel slopes for the Co based materials (Fig. 1f). The values between 70 and 38 mV/dec suggest a mixture between a reaction rate limited by the discharge reaction of protons at the surface (128 mV/dec, Volmer mechanism) and the Volmer-Tafel mechanism (38 mV/dec) where the recombination of adsorbed hydrogen is rate determining, which is observed after 1000 cycles on PdCoO$_2$. Pure Pd metal, which has a high Tafel slope of >100 mV/decade for current densities >1 mA/cm², is limited by the Volmer mechanism, in contrast to the delafossites presented herein, see Fig. S2. [42, 49, 50] The observed evolution of the activity described by $\eta_{10}$ in the delafossite oxides further appears to be related to the activity of the catalysts per surface area, $j_0$, (Fig. 1g), which changes as a function of time and parallels the trends observed for the overpotentials shown in Fig. 1 e. While $j_0$ tends to slightly decrease for PdCrO$_2$, it increases for both Co containing materials, indicating an effective increase in the intrinsic activity or catalytically active area, or both. The values for $j_0$ evolving with time are all in the range of 1 mA/cm² and thus amongst the highest reported to date. For PdCoO$_2$, $j_0$ of up to 5(1.2) mA/cm² is even superior to those observed for both bulk Pt and Pd (2.3 and 0.9 mA/cm², respectively), as shown in Fig. S2.[42, 43, 51]

In order to obtain an estimate of the real surface area, impedance measurements were performed to extract the double layer capacitance $C_{DL}$, which is proportional to the electrochemical surface area (ECSA), and the pseudocapacitive contribution $C_P$ (see discussion in the SI, section 3 and Fig. S3). Since due to their modified surface properties no references for the *operando* modified delafossites are available, the same capacitive response factors as for bulk Pd or Pt were assumed (see SI section 3 for details). We then obtain an ECSA normalized $j_{0,ECSA}$ of 0.14 mA/cm² for PdCrO$_2$ and 0.20 mA/cm² for (Pd,Pt)CoO$_2$ after 300 cycles (see Fig. S3 for details). These values are close to dealloyed Pd nanoparticles without normalization for real surface area (0.18 mA/cm²)[52] and reported surface normalized values of 0.14 mA/cm² for tensile strained Pd grown on Au or rough Pd surfaces (0.22 mA/cm²).[53, 54]

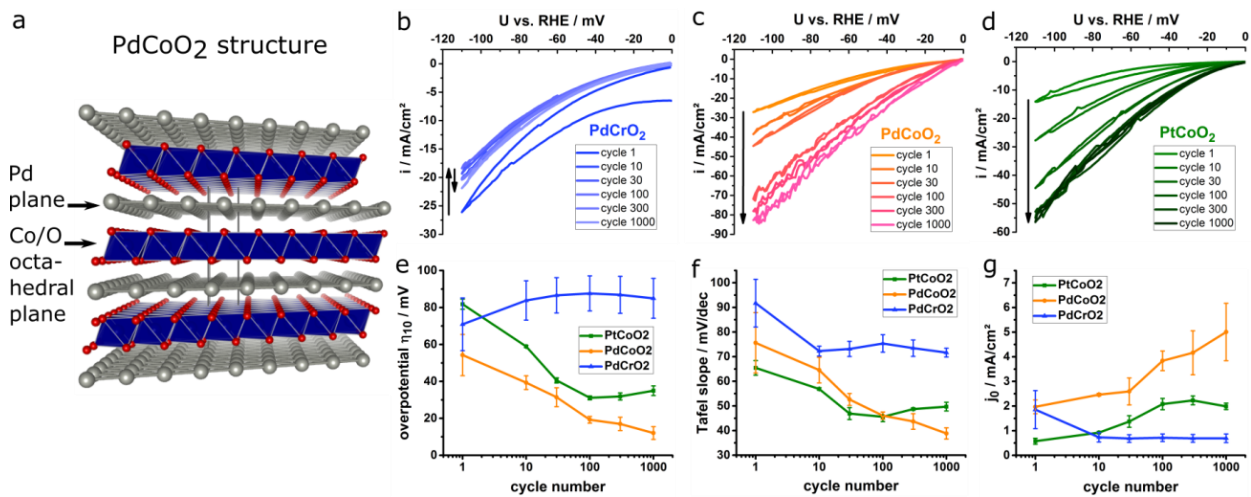

**Figure 1**: Crystal structure of the delafossites and the evolution of the electrochemical activity for HER. a: The layered delafossite structure using PdCoO$_2$ as a representative example, view along [100] (Pd: grey; oxygen: red, Co: blue octahedra). b-d: Uncorrected cathodic currents for PdCrO$_2$ (b), PdCoO$_2$ (c) and PtCoO$_2$ (d) in hydrogen saturated 1M H$_2$SO$_4$ during cathodic cycling (scan rate: 2 mV/s for displayed measurements, 25 mV/s for aging in-between the displaced cycles). d-f: The resulting

Tafel parameters (e: overpotential $\eta_{10}$, f: Tafel slope, and g: exchange current density for geometric surface area) after IR-correction and fitting in the region >1 mA/cm².

To provide insights into the origin of the currents observed in the electrocatalytic experiments, the evolution of hydrogen was investigated close to the surface by scanning electrochemical microscopy (SECM) measurements on $PdCoO_2$ single crystals *operando* (see Fig. S4 a & b for illustration). Although no site dependence could be mapped due to a tip-limited resolution >10 µm, the probing electrode feedback clearly shows the presence of hydrogen near the sample surface (~10 µm) when the sample is held at different cathodic sample potentials under weakly acidic conditions. Cyclic voltammograms reflecting the reductive SECM tip HER current for different sample potentials show a decrease when the $PdCoO_2$ sample HER current becomes competitive, and an increase in the hydrogen oxidation current at oxidative tip potentials (Fig. S4 c). Temporally, an increase in hydrogen production at the sample could be evidenced by increasing proton reduction currents at the tip (Fig. S4 d). This highlights the fact that pristine samples are HER active and improve over time, even without cycling. Furthermore, a comparison of the charge passed through a pre-aged $PdCoO_2$ single crystal (> 1000 cycles), referenced to a Pt wire, yielded stable Faradaic efficiencies towards hydrogen evolution of > 90%. In contrast, a pure Pd wire had a Faradaic efficiency of only 6%, probably due to large amounts of $H_2$ being incorporated into the Pd lattice, even after long coulometric measurements (see Fig. S5). [50, 51]

**Surface modification analysis**. Next, to understand the origin of the observed increase in catalytic activity over time, the modifications on the surface of the delafossites were tracked by analysis of the electrolyte composition by means of inductively coupled plasma optical emission spectrometry (ICP-OES) after the electrocatalytic experiments (see SI Table ST1). Almost no detectable amounts of Cr in the electrolyte suggest a high stability of this element in the $PdCrO_2$ surface (<0.08 at% of the Cr amount present in the electrode), while $PtCoO_2$ showed small amounts of Co in the electrolyte, corresponding to 0.24(3) at% of the electrode element. No dissolved Pt was evidenced in all cases. In contrast, $PdCoO_2$ shows increased amounts of Co in the electrolyte, indicating a much faster dissolution of Co, which progresses at a rate two orders of magnitude faster than Cr dissolution.[55] After 1000 cycles, 9.35(90) at% of the Co has leached out from $PdCoO_2$.

Consequently, we expect the formation of a Pd rich surface layer on $PdCoO_2$. Indeed, direct evidence of the surface transformation is obtained by optical microscopy and scanning electron microscopy (SEM) on the surfaces of the delafossites after catalysis. While $PdCrO_2$ does not show any obvious signs of corrosion (Fig. 2 a-b), the surface of $PtCoO_2$ is slightly modified, which is elucidated by a slight roughening and accumulation of heavier elements as shown by the back scattered electron (BSE) detector image (Fig. 2 c-d vs e-f). In contrast, the surface of $PdCoO_2$ is strongly modified, as shown in Fig. 2 g-l and mirrored by the significant Co dissolution and a stronger change in catalytic properties. Inspection under an optical microscope (Fig. 2 g-h) reveals a large area surface modification and roughening. A more detailed SEM analysis of the surface (Fig. 2 i vs j) shows the formation of a continuous capping layer with a higher concentration of heavy elements than the pristine, underlying material. Energy dispersive X-ray spectroscopy (EDS) maps on the modified surface further underline the enrichment of Pd in this topmost capping layer. The formation of cracks and nanoparticles within the capping layer (Fig. 2 j-l) suggests a release of strain still present in the remaining capping structure after Co and O are dissolved near the surface (*vide infra*). These cracks can also be related to potential dependent hydrogen loading and unloading into the catalysts during electrocatalytic cycling, which results in dynamically modified lattice parameters.[56] Also, small amounts of Pd were found in the electrolyte after catalysis, which can be explained by partial detachment of the strained capping layer (Fig. S6 a-c). [56]

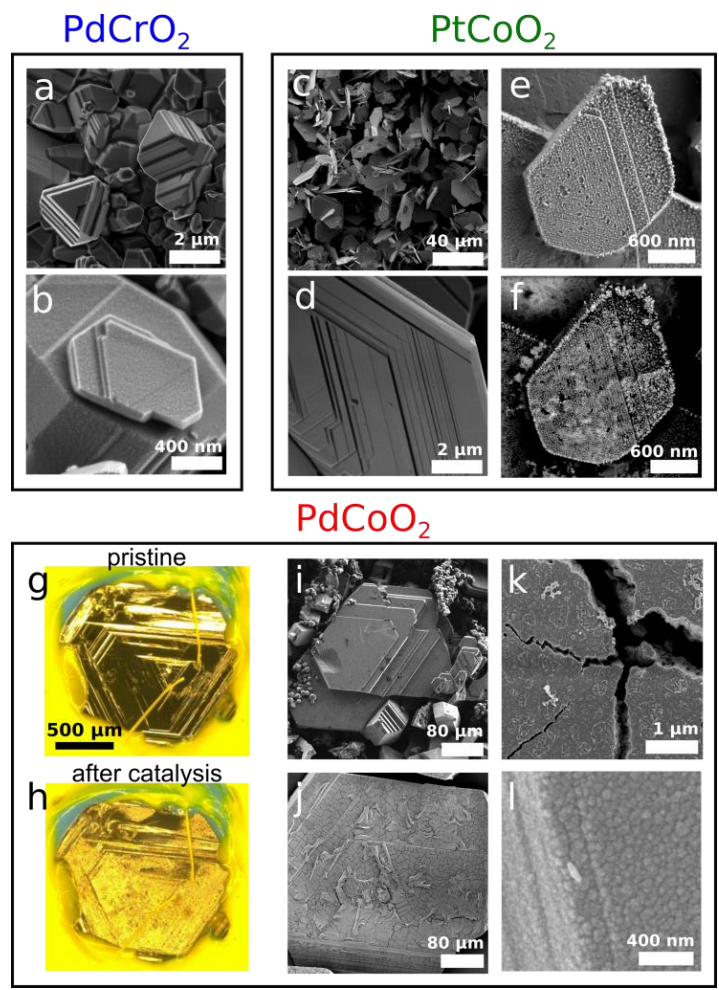

**Figure 2:** Surface modifications on the delafossites before and after electrocatalytic HER in 1M $H_2SO_4$. a: $PdCrO_2$ crystals as-synthesized. b: Enlarged image after cycling. c-d: As-synthesized $PtCoO_2$ crystals. e-f: PtCoO2 after cycling. The BSE detector image (f) highlights the heavy element nature of the capping layer, containing mostly Pt. g-h: Optical microscope images of a large $PdCoO_2$ crystal before and after catalysis, embedded in yellow sealing. i: SEM image of a large $PdCoO_2$ crystal embedded in a carbon paste electrode prior to cycling. j: $PdCoO_2$ surface modification after cycling showing the Pd rich capping layer. k: Zoom into the cracked surface structure. l: Magnification of the rough nanostructure on the surface.

To further investigate the surface modification, X-ray photoelectron spectroscopy (XPS) was used to measure the valence state of the respective components of all three materials before and after catalysis, as shown in Fig 3. A detailed discussion including survey spectra (Fig. S7) and peak lists (Tables ST2-5) can be found in the SI section 8. In brief, the main elemental lines of all pristine delafossites correspond well to the literature values. In the oxygen spectra (Fig. 3 c, f and i), an O 1s signal at 531-532 eV is attributed to surface adsorbed oxygen. After catalysis, the amplitude of the delafossite M-O interaction peak at 529 eV is slightly lowered relative to the surface O 1s peak in PdCrO$_2$ (Fig. 3c), suggesting somewhat decreasing binding of Pd and Cr to O at the surface. For PdCoO$_2$ (Fig. 3 d-f) and PtCoO$_2$ (Fig. 3 g-i), the Co as well as the M-O 1s signals completely disappear after catalysis, with slightly slower kinetics for PtCoO$_2$ (see Fig. S8), underlining the formation of a Pd(0) or Pt(0) layer on the surface during catalysis, in agreement with the ICP measurements on Co dissolution. Summarizing these results, we can conclude that PdCrO$_2$ remains largely unchanged over the cycling experiments, while PtCoO$_2$ and PdCoO$_2$ gradually lose Co and are eventually reduced, forming a metallic capping layer of the respective noble metal.

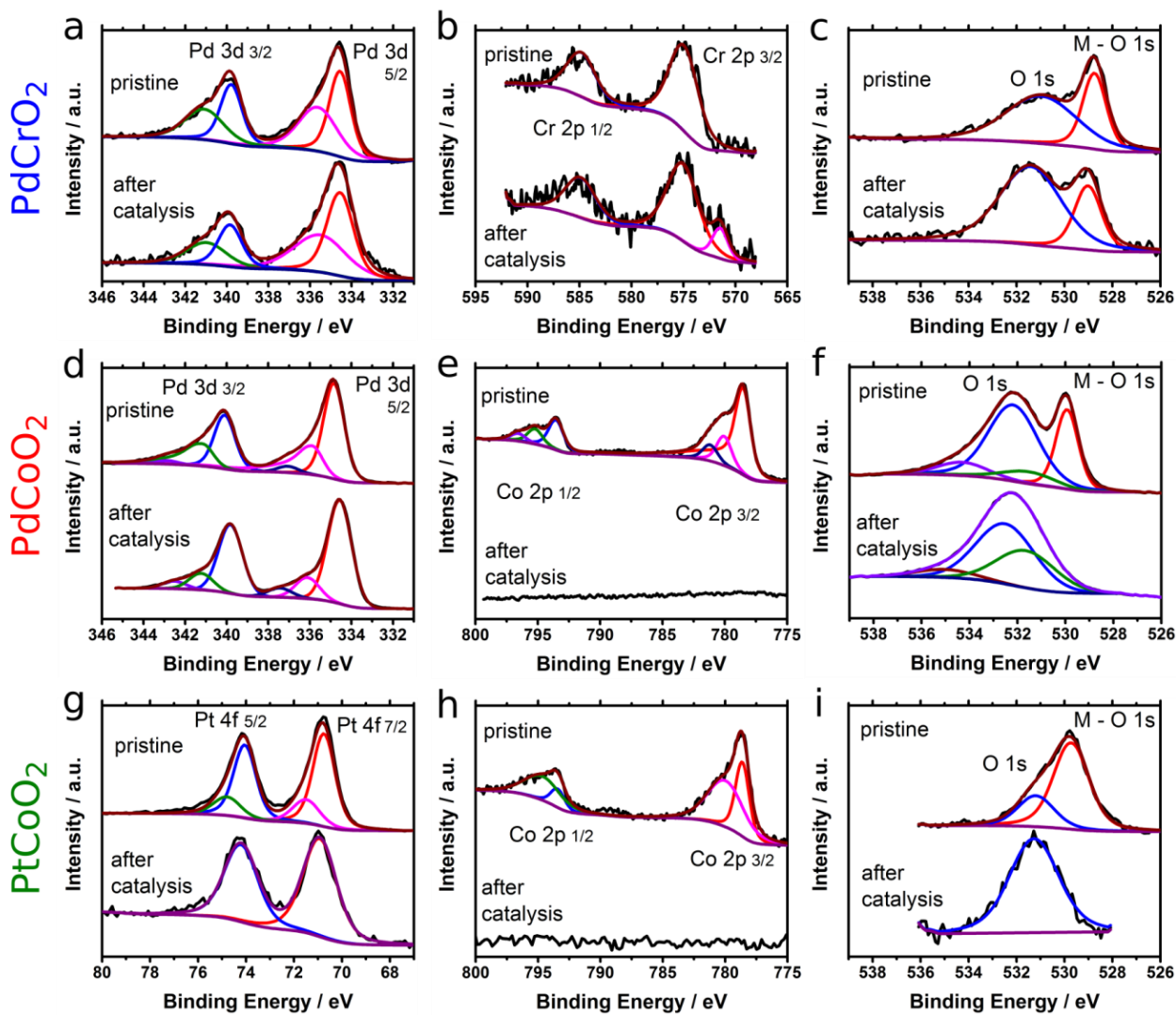

**Figure 3**: XPS spectra of PdCrO$_2$ (a-c), PdCoO$_2$ (d-f) and PtCoO$_2$ (g-i) before (top trace) and after (bottom trace) catalytic cycling in 1M H$_2$SO$_4$. While the surface of PdCrO$_2$ remains almost unchanged, the surface of PdCoO$_2$ is transformed into Pd with traces of oxygen adsorbed to the surface. For PtCoO$_2$, the behavior is similar and a Pt rich surface is obtained after longer aging. A peak list including all signals and their areas as well as the survey and additional spectra are shown in the SI section 8 (Fig. S7-8, Table ST 2-5).

**Capping structure and strain analysis.** Having identified the nature of the surface corrosion process as the reductive formation of a Pd rich capping layer on PdCoO$_2$, we turn to the question as to why this process leads to the observed exceptionally high activity towards HER. To this end, we first used scanning transmission electron microscopy (STEM) to characterize the structure and elemental composition of the pre- and post-catalysis samples at high spatial resolution. STEM lamellae were prepared from a PdCoO$_2$ single crystal after short (5 min) and long (hours) chronoamperometric aging at -100 mV vs. RHE, which both lead to the formation of a capping layer with increasing thickness (see Fig. 2 c-e and S6). A cross section lamella of a 5 min aged single crystal contains both the capping layer and the underlying material, as shown in Fig. 4 (see SI section 6 for experimental details and Fig. S 6 for a lamella prepared after hours of aging). The Pd capping layer is ca. 30 nm thick and consists of Pd nanocrystals of 2-10 nm size (Fig. 4 a-d). Elemental analysis of the capping layer by EDS spectrum imaging [57] shows that the cap consists primarily of Pd (Fig. 4 e), while the Co concentration is below the detection limit (< 2%). O is only observed at the surface of the Pd capping layer. The concentration profiles are summarized in the right display element of Fig. 4 f. As shown in Fig. 4 a, the Pd *fcc* capping layer grows directly on the PdCoO$_2$ surface, in direct contact to the electrolyte. There is a sharp transition between the PdCoO$_2$ crystal and the Pd cap (Fig. 4 b). The atomic resolution STEM-high angle annular dark field (HAADF) image of PdCoO$_2$ (Fig. 4 c) is overlaid by the crystal structure along the [11-20] zone axis. We observe a preferred orientation of the cap on the PdCoO$_2$ substrate, suggesting quasi-epitaxial growth of the Pd (111) in the c-direction of the bulk PdCoO$_2$ (normal to the hexagonal Pd sublattice). Within the cap structure, twin variants of Pd nanocrystals are observed (Fig. 4 d and S9). Fast Fourier Transformations (FFT) of the respective cap (Pd *fcc*) and bulk (PdCoO$_2$) regions reveal only a few degrees tilt between Pd (111) and PdCoO$_2$ (0003), as shown in Fig. S9 (a and b). This lends further evidence that the capping layer grows via leaching of Co and O from the bulk material, while the hexagonal Pd lattice in PdCoO$_2$ evolves into an *fcc* Pd lattice with minimal structural reorganization, as illustrated in Fig. 4 g. To derive the lattice parameter of the Pd nanocrystals, STEM micrographs were calibrated using the lattice parameters of PdCoO$_2$ from XRD measurements. Compared to the sharp reflections of PdCoO$_2$, Pd reflections are broader along the radial and azimuthal axes, corresponding to a broader range of strain and rotation among Pd nanocrystals. Analysis of all Pd {111} reflections evidence a homogeneous lattice parameter of 3.98 Å on average, which proves that the Pd *fcc* capping layer shows isotropic tensile strain, even for both twin variants (Fig. S9 c). Further comparison shows a slightly smaller lattice parameter at the top 5 nm of the cap than the bottom part (Fig. S9 e-g), suggesting partial strain relaxation towards the surface, especially for the Pd(111) planes that are parallel to the PdCoO$_2$ surface (Fig. S8, d-g). This is expected for quasi-epitaxial growth and partial strain release due to cracks. Nevertheless, the lattice parameters close to the surface are still larger than those of bulk Pd crystals (3.89 Å) [58] or the ones reported for Pd nanoparticles with similar size (approx. 10 nm) that show no lattice dilatation (see also Fig. S10).[59] Only below 3 nm, a comparable lattice parameter has been reported for pure Pd nanoparticles deposited on carbon. [60] A lattice constant of 3.98 Å in the capping layer corresponds to Pd nearest neighbour distances of 2.815 Å in the Pd *fcc* structure, which is very close to 2.83 Å in the Pd sublattice of PdCoO$_2$ and significantly differs from unstrained Pd (2.75 Å). This further underlines the relevance of lattice strain in the substrate material, PdCoO$_2$, which is transduced to the capping layer.

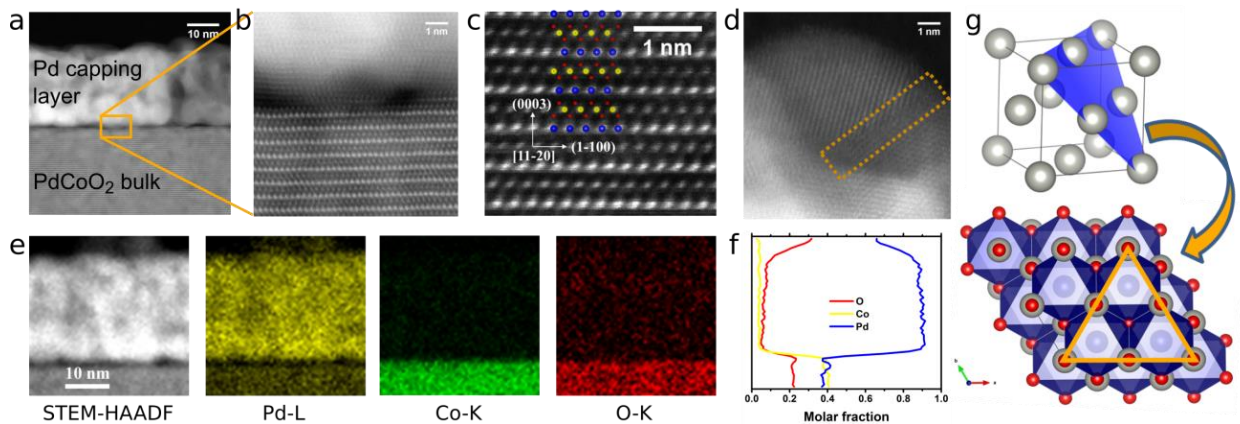

**Figure 4:** STEM analysis of the Pd capping layer on PdCoO$_2$. a: STEM-HAADF micrograph of PdCoO$_2$ after 5min aging at $\eta$ = 100 mV in 1M H$_2$SO$_4$, showing the bulk material and the Pd rich capping layer. b: PdCoO$_2$ atomic structure (Pd: blue, Co: yellow, O: red). c: Pd *fcc* nanocrystals in the overlayer containing twinned grain boundaries (marked in orange). d: STEM-HAADF image and respective STEM-EDS elemental maps based on the Pd-L line (blue), the Co-K line (yellow) and O-K line (red) and a line profile of their molar fractions (right element). e: Illustration of the Pd *fcc* structure with the (111) plane corresponding to the hexagonal Pd sublayers in PdCoO$_2$, which act as a growth template for Pd *fcc*.

The observations made on PdCoO$_2$ in Fig. 4 agree with ICP measurements of dissolved Co in the electrolyte and the XPS data (Fig. 3) but further suggest that the continuous dissolution of Co and O creates a significantly strained Pd-rich capping layer, where the strain is transmitted by the pseudomorphic growth of Pd(111) on the hexagonal Pd sublattice in PdCoO$_2$.

**Pd nanoparticle reference measurements**. Since the particle size and density of surface states of Pd nanocrystals in the capping layer (2-10 nm) differs from that of bulk Pd, which can affect the catalytic activity,[59, 61] we studied the electrochemical activity of Pd nanoparticles of comparable size (1 – 10 nm, *Sciventions*) to gauge the effect of the particle morphology and strain on the catalytic activity (see Fig. S10-11). Rietveld refined XRD measurements of the dried reference nanoparticles revealed an average lattice constant of 3.90(2) Å (see Fig. S10), which fits reported values for bulk Pd (3.89 Å) and Pd nanostructures of 8-10 nm size.[59] The lattice constant of the nanoparticles is thus 2% smaller than for the nanocrystals at the bottom of the Pd capping layer (3.98 Å). Since a comparable surface coverage and active surface area is difficult to achieve, the catalytic properties of both bulk Pd and the nanoparticles are best characterized by their Tafel slope, which is only sensitive to the underlying catalytic process and independent of the actual effective surface area.[42] Both the nanoparticles and the bulk Pd show a similar Tafel slope of ~105 mV/dec (Fig. S11 and S2), which is consistent with the Pd literature value of 103 mV/dec [49] and relates the surface catalytic process to a rate limiting Volmer (discharge) reaction. This value is much larger than the values found for surface modified PdCoO$_2$ in Fig. 1 and S2 (38 to 30 mV/dec), which suggests a rate limiting Volmer-Heyrovsky or -Tafel mechanism (38 and 30mV/dec, respectively). [42, 62] The intrinsic catalytic mechanism of the strained Pd capping layer is thus different from pure Pd, regardless of particle size or morphology.

**Discussion.** To rationalize the observed tensile strain-induced activity increase in $PdCoO_2$, we first consider the effect of strain as expected from classical d-band theory, which predicts a correlation between the position of the metal d-band center with the HER activity.[53] The relative energies of the metal d-band and the antibonding hydrogen $\sigma^*$-orbital directly determine their overlap and, hence, bonding strength of the H-adsorbate to the metal surface, $\Delta G_H$.[63] Following Sabatier's principle, the catalyst – substrate interaction should be neither too strong nor too weak such that both adsorption of the substrate and desorption of the product is facile. Hence, $\Delta G_H$ is used as the most relevant descriptor of the intrinsic activity for HER, given by a turnover frequency or $j_0$, resulting in a volcano plot, with optimal catalyst – substrate interactions at the top of the volcano.[42, 64] Tensile lattice strain in Pd tends to flatten out the d-bands while upshifting the band center towards the Fermi level, leading to stronger overlap with the hydrogen $\sigma^*$-orbital and, thus, stronger metal—H bonds.[12, 65] In fact, growth of a pseudomorphic Pd layer on a single crystalline substrate with a larger lattice constant has been shown to increase $\Delta G_H$ and thus move tensile strained Pd away from the top of the volcano while decreasing the exchange current density.[13, 53] Interestingly, we observe the opposite trend, enhancing both $j_0$ and reducing the Tafel slope. The observed change in Tafel slope by a factor of almost 3 with respect to bulk Pd suggests a more fundamental change in the materials surface properties to be at play. Indeed, tensile strain positively affects the hydrogen adsorption capacity of Pd and has been described to modify the potential of hydrogen adsorption in Pd.[53, 66, 67, 68] On the other hand, the different lattice parameters between hydrogen poor α-Pd and the hydrogen rich β-phase (sharp transition and discontinuous lattice parameter transition from 3.89 to 4.02 Å) kinetically hinder a phase transformation.[69, 70, 71] Since our expanded Pd *fcc* lattice intrinsically shows stable lattice parameters (as probed by TEM after prolonged ex situ sample storage) close to those of β-PdH$_x$ (3.98 Å and 4.02 Å, respectively),[71, 72] incorporation of hydrogen is expected to be more facile.[66] This conjecture rationalizes the corresponding findings on the Tafel slopes, which indeed show β-PdH$_x$-like behavior (30-38 mV/dec Tafel slope). [51, 62] To investigate the possible *operando* formation of a hydride phase, faster cyclic voltammetry measurements (25 mV/sec and 100 mV/sec) were recorded after 1000 cycles (Fig. S 12). With increasing scan rate, a stronger capacitive contribution is observed. It is most prominent for PdCoO$_2$, as evidenced by the separation of the forward and backward scans around the open circuit potential (OCP), and the OCP shifts to more cathodic potentials for this material (Fig. S12 b, 100mV/s scans). This indicates, in accordance with the impedance data presented in Fig. S3, that the capping layer is active for hydrogen adsorption - a well-known phenomenon in Pd metal that forms interstitial hydrides (PdH$_x$). [48, 50, 51] Since the sorption effects are very fast (~1 s), the transition to β-PdH$_x$ can occur almost immediately once a strained Pd capping layer has been formed under reductive conditions.

In summary, we have investigated the delafossite oxides PdCrO$_2$, PdCoO$_2$, and PtCoO$_2$ for HER activity in acidic media for the first time. In their pristine form these materials outperform most reported electrocatalysts with respect to their overpotentials required for 10 mA/cm²$_{geo}$ (71(15) mV for PdCrO$_2$, 54(12) mV for PdCoO$_2$, 82(3) mV for PtCoO$_2$ in the first cycles). The very high activity for HER originates from a superior intrinsic surface activity, translating into exchange current densities on the order of mA/cm² (initially 1.9(8) mA/cm² for PdCrO$_2$, 2.0(3) mA/cm² for PdCoO$_2$, 0.57(12) mA/cm² for PtCoO$_2$), which are orders of magnitude higher than most HER catalysts besides Pt. This property is especially beneficial for HER at small overpotentials, where the effect of activity increase by the Tafel slope does not yet come into play. While PdCrO$_2$ is relatively stable under acid HER conditions, the surfaces of PtCoO$_2$ and especially PdCoO$_2$ corrode over time *via* the reductive dissolution of Co and O at the surface near region. The resulting Pd or Pt enriched capping layers show significantly enhanced activities compared to the bulk materials: $j_0$ increases by a factor of 2.5 and 3.5 for PdCoO$_2$ and PtCoO$_2$, while the Tafel slopes decrease from 76(13) to 38(3) mV/dec and from 65(3) to 50(2) mV/dec, respectively.

In the case of PdCoO$_2$, the tensile strain, which is present in the noble metal sublattice of all these delafossites, is translated to the Pd *fcc* capping layer, which grows directly on the underlying bulk material. This strain increases the hydrogen adsorption energy as predicted by d-band theory and lowers the energetic barrier for the formation of a β-PdH$_x$ phase, which otherwise requires a significant lattice expansion (+3.3%). This phase transformation entails a change in HER mechanism, shifting the surface catalytic process away from the rate limiting Volmer-reaction found in Pd towards the more efficient Volmer-Tafel or Volmer-Heyrovsky mechanism. The resulting material has an overpotential of 12(3) mV for 10 mA/cm²$_{geo}$ only, hence outperforming even bulk Pt.

The herein presented strain engineering points towards a more general design principle for the rational activity enhancement in electrocatalysts with inherently strained metal sublattices. It further suggests that phase transformations, stabilized by strain, may be key players in determining both metal—substrate interactions and the intrinsic catalytic activity, and as such, strain-induced phase changes may complement classical d-band theory as a model to predict catalytic performance.

## Acknowledgements

We gratefully acknowledge Peter Schützendübe and Michaela Wieland for XPS measurements, Eleonora Frau and Pranit Iyengar for the introduction to and assistance with SECM measurements. Y. Eren Suyolcu, Aliaksandr Bandarenka and especially Rotraut Merkle are acknowledged for fruitful discussions. E.A.L. acknowledges the support from the SNF Ambizione Energy program and the research program of FOM, which is financially supported by The Netherlands Organization for Scientific Research (NWO). S.Z. and C.S. acknowledge financial support from the German Research Foundation DFG under the priority programme SPP 1613 [DFG SCHE 634/12-2]. F.P, E.A.L, B.V.L and A.F.M thank the MPS-EPFL center for financial and logistic support.

## Data availability

The data supporting the plots within this paper and other findings of this study are available from the corresponding author upon reasonable request.

## Author contributions

F.P., D.W., F.H. and B.V.L conceived the project and the contributing measurements. The materials were synthesized by D.W., L.D. and R.E. All sample preparation and electrochemical measurements were done by F.P. The SECM data was analyzed and discussed by E.A.L and F.P. G.R. and F.P. analyzed the XPS data. S.Z. performed the STEM experiments including the data analysis and presentation. F.P. created all other graphs. F.P. and B.V.L wrote the manuscript. All authors including C.S. and A.F.M. contributed to the discussion of the measurements, the data interpretation and the manuscript discussion.

## Conflict of interest

The authors declare no conflict of interest.

71. Ulvestad A, Welland MJ, Collins SSE, Harder R, Maxey E, Wingert J*, et al.* Avalanching strain dynamics during the hydriding phase transformation in individual palladium nanoparticles. *Nature Communications* 2015, **6:** 10092.

72. Akiba H, Kofu M, Kobayashi H, Kitagawa H, Ikeda K, Otomo T*, et al.* Nanometer-Size Effect on Hydrogen Sites in Palladium Lattice. *Journal of the American Chemical Society* 2016, **138**(32)**:** 10238-10243.
27